\documentclass{aa}
\usepackage{graphicx}

\newcommand{\ms}{M$_{\sun}$ }
\begin{document}                                          
\title{The surprising Far-UV spectrum of the polar BY\,Camelopardalis
\thanks{This work is based on data obtained from the NASA-CNES-CSA FUSE 
mission}}
\author{M. Mouchet\inst{1,2}
\and J.-M. Bonnet-Bidaud\inst{3}
\and E. Roueff\inst{1}
\and K. Beuermann\inst{4}
\and D. de Martino\inst{5}
\and J.M. Desert\inst{6}
\and R. Ferlet\inst{6}
\and R.E. Fried\inst{7}
\and B.T. G\"ansicke\inst{8}
\and S.B. Howell\inst{9}
\and K. Mukai\inst{10}
\and D. Porquet\inst{11}
\and P. Szkody\inst{12}}
\offprints{M. Mouchet, email:  martine.mouchet@obspm.fr }

\institute
{UMR 8102, associ\'ee au CNRS et \`a l'Université Paris 7,  
LUTH, Observatoire de Paris, Section de Meudon, F-92195 Meudon Cedex, France
\and
Universit\'e Paris 7 - Denis Diderot, 2 Place Jussieu, F-75005 Paris, France
\and
Service d'Astrophysique, DSM/DAPNIA/SAp, CE Saclay, 
F-91191 Gif sur Yvette Cedex, France
\and
Universt\"ats-Sternwarte G\"ottingen, Geismarlandstr. 11, 37083
 G\"ottingen, Germany 
\and INAF, Osservatorio Astronomico di Capodimonte, Via Moiariello 16, 80131 
Napoli, Italy
\and UMR 7095,  associ\'ee au CNRS et \`a l'Université Paris 6, 
Institut d'Astrophysique de Paris, 98 bis Boulevard Arago, 
F-75014 Paris,  France 
\and Braeside Observatory, Flagstaff, AZ 86002, USA
\and Department of Physics and Astronomy, University of Southampton, 
Southampton SO17 1BJ, UK
\and Astrophysics Group, Planetary Science Institute, Tucson, AZ 85705, USA
\and NASA/Goddard Space Flight Center, Greenbelt, MD 20771, USA
\and Max-Planck-Institut f\"ur extraterrestrische Physik, D-85741 Garching, 
Germany 
\and Department of Astronomy, University of Washington, Seattle, WA 98195, USA
}
\date{Received: 2 October 2002/ Accepted: 31 January 2003}
\titlerunning{The far-UV spectrum of BY Cam}
\authorrunning{Mouchet et al.}

\abstract{
We report on the first far-UV observations of the asynchronous polar BY Cam 
made by the {\it Far-Ultraviolet Spectroscopic Explorer (FUSE)}.
The source is known to exhibit the most extreme NV/CIV emission resonance 
line ratio observed among polars. 
The FUSE observations reveal a  OVI resonance line weaker 
than in the prototype of polars, AM Her, with the absence of a detectable 
narrow component. The OVI broad line is detected  with an 
equivalent width of the same order as in  AM Her, the blueward doublet
 component is clearly present but the redward 
component is strongly affected by H$_2$ absorption. 
The presence of a strong NIII line and weak CIII lines also
confirms the peculiar CNO line flux. We compare the resonance CNO line 
intensities with the predictions of the CLOUDY plasma code coupled 
to a geometrical model of the accretion column. 
Varying the temperature and/or intensity of the
ionising spectrum is unable to reproduce the observed broad line ratios.
A solution is obtained by significantly  altering the element abundances with  
a strong depletion of C, overabundance of N  and a weak underabundance of O.
This confirms previous suggestions of non-solar abundances which may result 
from redistribution in the accreted material following nova outbursts and/or 
the secondary nuclear evolution.
A very significant H$_2$ absorption is observed in front of the source, a 
possible indication for either the existence of a dense interstellar cloud
 or of circumstellar material.
 } 
\maketitle

\section{Introduction}
The hard X-ray source H0538+68/BY Cam was identified as a polar,  
a cataclysmic binary with a highly magnetized  
(B\,$\sim 10^7-10^8$ G) accreting white dwarf, after the detection of 
circularly polarized optical flux by Remillard et al. (1986). 
It was shown to exhibit a very peculiar UV emission line 
spectrum, with an emission resonance 
line ratio NV/CIV, more than ten times larger than what is usually 
observed in Polars (Bonnet-Bidaud \& Mouchet 1987 (BM87)). 
It has been suggested that this unusual feature could be the result of 
a chemical evolution 
of the secondary, or of an unnoticed nova event leading to a 
CNO redistribution (BM87, Mouchet et al. 1990a). 
The discovery that the source is  slightly desynchronised 
(Silber et al. 1992) and the identification of a known nova, V1500 Cygni, 
as a polar which is also desynchronised (Schmidt \& Stockman 1991), gave 
further arguments to the nova hypothesis. 
However the picture has gradually changed as at least two synchronous polars, 
V1309 Ori (Szkody \& Silber 1996) and MN Hya (Schmidt 
\& Stockman 2001), have now been found to show a large NV/CIV ratio
though less extreme than in BY Cam.
The origin of these anomalous line strengths among polars is not clear. 
It might obviously result from non-solar abundances in the accreted matter
but could also be due to peculiar ionisation conditions. These effects
may strongly affect our understanding of the evolution of these systems and 
their link to  novae.

The FUSE satellite, operating in the range 905-1187\AA{} offers 
 access to the OVI resonance line, a key for evaluating the CNO abundances.
We report here on the first observation of BY Cam by this satellite. 
These observations were simultaneous  with RXTE and EUVE observations 
as well as optical spectroscopy with the 3.5m APO telescope and with the 9m 
HET, and photometry at the 16'' Braeside telescope whose results, including 
a temporal analysis, will be reported elsewhere.

\section{FUSE observations and reduction}
BY Cam was observed on 2000 January 10 and 16 by the FUSE satellite 
(Moos et al. 2000, Sahnow et al. 2000). The total wavelength range of 
905-1187\AA{} is covered  by means of four gratings and mirrors leading to
bandpasses of about 200\AA{}, in such a way that the region below 
$\sim$ 1000\AA{}
is covered by two detectors, the central region by four  and the 
region above $\sim$ 1100\AA{} by two.    
During the first observation, the source was badly centered  
in the aperture and only the Jan. 16 data are presented  below.
The dataset consists of five exposures in the large aperture (30''x30''), 
with exposure times ranging from 3.44 to 5.42\,ks, resulting in
a total 22\,ks on-source time. This covers two orbital cycles 
with only a small orbital phase interval ($\sim$0.12) not covered.
The source was in the satellite continuous viewing zone and observations 
occurred mainly during 'day' time, leading to the presence of strong 
geocoronal lines.

The data have been reduced using the Calfuse 1.7.3 version of the 
standard pipeline.  For each detector the five exposures were co-added  
with an eventual wavelength shift correction of one pixel. 
Velocity measurements have been corrected for the error in 
the heliocentric velocity value present in this
calibration pipeline version. 
As the source is faint, the resulting signal-to-noise ratio per resolution 
element is quite low (less than 3) so the original (0.005\AA) data have 
been binned into 0.1\AA\, intervals. 
We report here on the mean phase-averaged spectrum.

\section{The average spectrum}
The average spectrum obtained by combining all detectors and exposures
is shown in Fig. 1 where the strongest geocoronal lines have been 
removed for clarity. Some regions, such as around the NIV/HeII/H and HeII 
lines, are however possibly still contaminated by airglow
and should be treated with care.

\begin{figure}
\resizebox{\hsize}{!}{\includegraphics[angle=-90, bb=0 50 570 690]{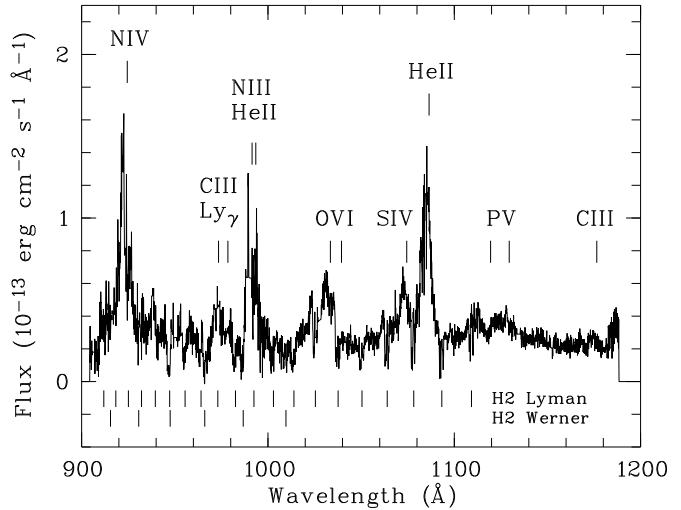}}
\caption{ The FUSE phase-averaged spectrum of BY Cam. The geocoronal lines 
have been removed for clarity and data are shown in 0.1\AA\, bins.
Remaining geocoronal lines can still contaminated  the feature near 920\AA.
The tick marks at the bottom  indicate the positions of the R(0)  H$_2$ 
absorption lines of the lowest vibrational level v''=0 for the Lyman and 
Werner bands. }
\label{fig1}
\end{figure}

The continuum energy distribution is consistent with an extrapolation of
the IUE high state spectrum. 
It can be described with a energy power law  
F$_{\lambda} \propto \lambda^{-1.43}$ 
if no reddening is assumed (-2.44 for E$_{\rm B-V}$ = 0.05 (BM87)) with 
a total luminosity L(910-1185\AA) of $6.3\times 10^{31}$  erg\,s$^{-1}$ 
($12.2\times 10^{31}$  erg\,s$^{-1}$ for E$_{\rm B-V}$ = 0.05), 
assuming the source at 250 pc (G\"ansicke 1997)
and a Cardelli extinction law 
(Cardelli et al. 1989) with a standard A$_{\rm V}$/E$_{\rm B-V}$ = 3.1 value.
Broad emission lines are identified in Fig. 1, with intensities 
listed in Table 1.
Pronounced H$_2$ molecular lines of the Lyman and Werner series are 
also clearly present (see Sect. 4.1).   

\begin{figure}
\resizebox{\hsize}{!}{\includegraphics[angle=-90,bb=162 70 568 514,clip]{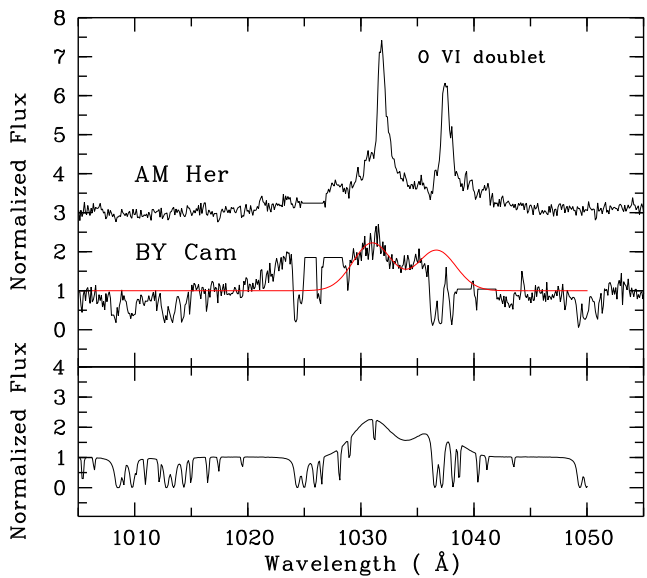}}
\caption{
Upper panel : the O VI doublet region as observed for AM Her (top, ORFEUS) and 
BY Cam (bottom, FUSE) with a best fit of two gaussians superposed. The 
geocoronal 
lines, inluding Ly$\beta$, are cut for clarity.
Lower panel : the two gaussian best fit including the H$_2$ absorption 
(see text).
}
\label{fig2}
\end{figure}

\bigskip
\begin{table}
\caption[ ]{FUSE emission lines}
\begin{flushleft}
\begin{tabular}{lcll}
\hline
\multicolumn{1}{c}{Line}    & \multicolumn{1}{c}{$\lambda$ (\AA)} & 
\multicolumn{1}{c}{Flux(1)} & 
\multicolumn{1}{c}{EW (\AA)} \\
\hline
N IV/HeII/H	&  923	& 4.8(+)  & 14.5	\\
CIII        &  977	& 0.6(+)  & 2.7   \\
N III/ HeII	&  989-92	& 2.6     & 10.2	\\
O VI		& 1032-38	& 2.0	    & 8.2   \\
S IV		& 1073	& 0.8	    & 2.1	\\
He II		& 1085	& 4.1(+)  & 11.0	\\
C III		& 1175	& 0.1	    & 0.6	\\
\hline
\end{tabular}
\end{flushleft}
(1) Line flux in units of \,10$^{-13}$ erg\,cm$^{-2}$\,s$^{-1}$\\
not corrected for reddening and/or N$_{\rm H_2}$ \\
(+) possibly contaminated by airglow
\end{table}

\begin{table}
\caption[ ]{Intensities and EW of the OVI doublet}
\begin{flushleft}
\begin{tabular}{lrrrr}
\hline
\multicolumn{1}{c}{} & \multicolumn{2}{c}{Narrow} & 
\multicolumn{2}{c}{Broad}  \\
& $\lambda$1032 & $\lambda$1038  & $\lambda$1032 & $\lambda$1038 \\ 
\hline
 & & & & \\
BY Cam Flux (*) & $\leq$0.04   &  -        & 1.4  & 1.1   \\
AM Her Flux (*) & 6.6       & 5.9 & 12.8  & 7.7  \\
\hline
BY Cam EW (\AA) & $\leq$0.15   &  -     & 5.2   & 4.4  \\
AM Her EW (\AA) & 2.3         & 2.1  & 4.6 & 2.8 \\
\hline
\end{tabular}
\end{flushleft}
(*) Line flux in units of \,10$^{-13}$ erg\,cm$^{-2}$\,s$^{-1}$\\
For AM Her : intensities derived from the mean ORFEUS spectrum (Mauche \& Raymond (1998))\\
For BY Cam: intensities derived assuming a FWHM of 0.5\AA\, 
(narrow) and  4\AA\, (broad) and corrected for H$_2$ molecular absorption. 
\end{table}

Figure 2 shows an enlargement of the OVI line region. The OVI doublet in BY Cam
is rather broad and weak, with the OVI $\lambda$1038 component affected 
by strong H$_2$ absorption. 
For comparison, the phase-averaged ORFEUS spectrum of AM Her 
(Mauche \& Raymond 1998), 
computed from retrieved archived data, is also shown in Fig. 2. 
Contrary to AM Her (see also Hutchings et al. 2002), 
no narrow OVI component is detected in BY Cam. 

The OVI broad line intensities have been derived by fitting two gaussians of 
same widths (FWHM of 1200 km\,s$^{-1}$), 
at the doublet separation, convolved with an absorption with a H$_{2}$ column 
density of 3$\times 10^{19}$ molecules\,cm$^{-2}$ (cf. Sect. 4.1). 
The results are given in Table 2 and the best fit is shown in Fig. 2 
(lower panel). 
Uncertainties in the doublet flux measurements have been computed by varying
the amplitude of each gaussian around the best value. They are given for
 the 99\% confidence 
level, corresponding to a $\chi^2$ variation of 6.6 for a 
one parameter-fit.  
Line intensities corrected for H$_2$ absorption and corresponding errors are 
$1.40^{+0.13}_{-0.12}$ and $1.09^{+0.22}_{-0.17}$ \,10$^{-13}$ erg\,cm$^{-2}$\,s$^{-1}$,
respectively for $\lambda$1032 and $\lambda$1038.
 Some caution should however be taken in interpreting the redward 
component intensity since it is affected by a strong H2 absorption 
and its red wing is also contaminated by an OI airglow contribution.
For AM Her, the OVI total (broad+narrow) equivalent width is larger than
for BY Cam but  the equivalent width of the broad component alone is of 
the same order.
The doublet intensity ratio of the broad component is also significantly 
different with a ratio of $\sim$1.3 for BY Cam and 1.7 for AM Her.
Assuming the source at 250 pc, the OVI broad line luminosity is 
$3.4\times 10^{30}$ erg\,s$^{-1}$, after correction for a E$_{\rm B-V}$ = 0.05
interstellar reddening. 
This can be compared to the dereddened NV($\lambda$1240) and 
CIV($\lambda$1549) luminosities
of $1.7\times 10^{31}$ erg\,s$^{-1}$ and $2.9\times 10^{30}$ erg\,s$^{-1}$ 
respectively,  derived from IUE observations (Zucker et al. 1995, 
Mouchet et al. 2002).
The mean ratios NV/CIV and OVI/CIV are 5.7 and 1.14 for BY Cam,  
while they are equal to  0.12 and 0.34 for AM Her. 
For this source, the values are derived from HUT spectra which wavelength
range includes the three resonance lines 
(Greeley et al. 1999) and refer to the broad component only. 
Note however that for BY Cam the 
values obtained at different epochs are also affected by a mean intrinsic 
$\sim$35\%  variability (Mouchet et al. 1990b, Zucker et al. 1995).

Contrary to AM Her, no strong narrow OVI line is observed in BY Cam, 
as shown in Fig. 2, where spectra for both sources are orbitally phase 
averaged. 
For BY Cam, the radial velocity amplitude of $\sim$\,250\,km\,s$^{-1}$ 
as derived for the 
optical narrow components (Mouchet et al. 1997), will introduce a broadening
but will not be sufficient to smear out a narrow component (FWHM $\sim$0.5\AA)
as intense as in AM Her. 
The absence of narrow lines in BY Cam may be due to the
absence of a strong soft X-ray component in BY Cam. However the polar VV Pup
which is a soft X-ray source, does not 
 either show obvious narrow components in the OVI lines 
(Hoard et al. 2002). 
Note that no conclusion 
can be drawn about the presence of a narrow component in the NV and CIV 
lines of BY Cam  due to the lack of high resolution UV spectra.

\begin{figure}
\resizebox{\hsize}{!}{\includegraphics[angle=-90,bb=25 50 580 590]{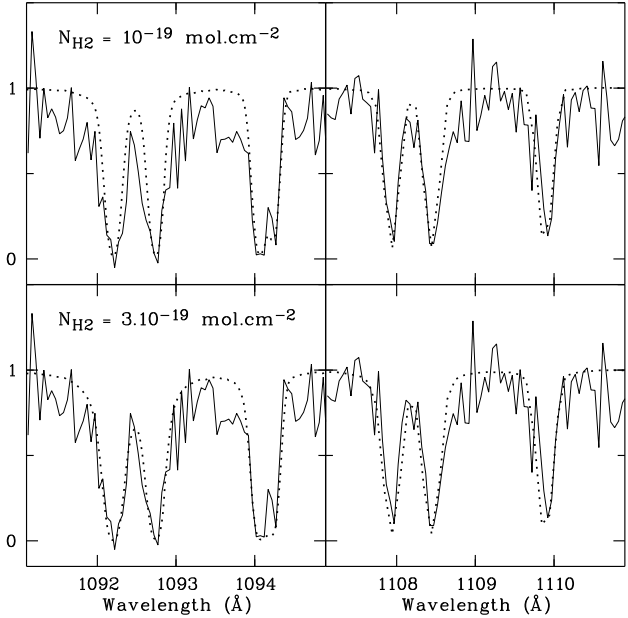}}
\caption{ Regions of the BY Cam normalized spectrum around the H$_2$ 
absorption lines
corresponding to the 0-0 and 1-0 vibrational transitions and models (dotted 
lines) for 
the best fit solution with N$_{\rm H_2}$ = 3$\times 10^{19}$ mol.\,cm$^{-2}$ 
(lower panels) and for  N$_{\rm H_2}$ = 1$\times 10^{19}$ mol.\,cm$^{-2}$
(upper panels) for illustration of the accuracy.
}
\label{fig3}
\end{figure}

\section{Discussion}
\subsection{The column density of molecular hydrogen}
Several H$_2$ absorption lines of the Lyman and Werner electronic system 
bands are clearly detected in the average FUSE spectrum of BY Cam. 
The spectrum of molecular hydrogen was modelled assuming specific values of the
total column density, a temperature of 80K and a b Doppler factor of 
10\,km\,s$^{-1}$.
The wavelengths, oscillator strengths and total radiative lifetimes of the 
upper levels of H$_2$ transitions are given by Abgrall et al. (2000).
The signal-to-noise ratio is too low to perform a fitting
of the complete spectrum as done in FUSE studies towards translucent 
interstellar  clouds (cf. Ferlet et al. 2000). We have compared 
the equivalent width measurements of the strongest  H$_2$  lines 
absorbed from the J=0, 1, and 2 levels in the v'-0 Lyman bands 
with v'\,=\,0 to 4, with those measured in the models.   
A best fit value is obtained 
for a H$_2$ column density N$_{\rm H_2}$ of  $3\times 10^{19}$ 
molecules\,cm$^{-2}$. 
The model spectra for the Lyman (0,0) and (1,0) bands are displayed 
in Fig. 3 for 2 different values of the H$_2$
 column density of 1 and 3$\times 10^{19}$ molecules\,cm$^{-2}$ together with 
the observed spectrum in the relevant spectral windows. 
The derived  H$_2$ column density of $3\times 10^{19}$ molecules\,cm$^{-2}$ 
  should be correct within a factor of 2.
When using the R(0) and R(1) lines only, a model with a lower b value of 
3 km\,s$^{-1}$  is marginally favoured.   
The presence of the R(3) component at 1067.6\AA{} and possibly at 1054.0\AA{}
seems to indicate a higher excitation temperature, but the quality of the 
spectra does not allow a reliable determination of its value.
The individual spectra are also relatively noisy 
so that it is not possible to check reliably for possible changes with 
orbital phase.\\    
From the absorption bump around 2200 \AA\,, an upper limit of 
E$_{\rm B-V}\leq$ 0.05 has been derived for the source (BM87), which 
corresponds to a neutral H density 
N$_{\rm H_1} ~\leq ~2.6\times 10^{20}$ cm$^{-2}$ (Shull \& Van Steenberg 1985).
This value is also consistent with the equivalent neutral H column density
N$_{\rm H_1} \sim (1-2)\times 10^{20}$ cm$^{-2}$ derived from
the absorption cut-off in the X-ray spectra (Kallman et al. 1993, 1996).
Taken at face value, this implies a very high 
N$_{\rm H_2}$/N$_{\rm H_1}$ ratio in the 
direction of the source. 
Assuming a standard interstellar medium, the expected
N$_{\rm H_2}$ value, consistent with the observed reddening, is only in 
the range 
$5\times 10^{11} - 5\times 10^{17}$ cm$^{-2}$ (Savage et al. 1977). 
A large N$_{\rm H_2}$ value is unexpected for a relatively nearby source. 
It may result from an inhomogeneous interstellar medium with a 
molecular cloud in the line of sight
but may also be linked with dense matter close to the source, possibly 
ejected during a prior nova event. 
Note that such  H$_2$ absorption has also being detected in the FUSE 
spectra of the supersoft binary
QR And and has been suggested to come from a circumbinary location 
(Hutchings et al. 2001). 
The present FUSE absolute wavelength calibration is not accurate 
enough to derive precise values of the H$_2$ line velocities and the 
systemic velocity 
of BY Cam is also not well known (in the range 1.3\,-\,141 km\,s$^{-1}$, 
Mason et al. 1989, Silber et al. 1992) so that no direct conclusion can 
yet be drawn about the possible circumbinary nature of the H$_2$ cloud.

\subsection{Model of the OVI line and CNO abundances}
The broad resonance lines of the highly ionised species of CIV, NV, and 
OVI (FWHM $\sim$ 1200\,km\,s$^{-1}$) are expected to be formed mainly 
in the accreting column, photoionised by the X-ray flux. 
Different attempts have been made to reproduce the line
ratios using photoionisation models with a simplified geometry (cf.
Mauche et al. 1997).
We evaluate here the line luminosities computed using the 
CLOUDY plasma code (Ferland et al. 1998), coupled to a geometrical model
of the accretion column. 
The ionising spectrum was defined as the sum of a 20\,keV bremsstrahlung,
as observed in the  RXTE observations, and of a
50\,eV blackbody   with a bolometric luminosity  of 0.1 times the hard
X-ray luminosity (Ramsay et al. 1994).
We have modelled the accretion column assuming a dipole geometry and  
free-fall velocities along the accretion column
(cf. Stockman  \&  Schmidt 1996). 
The density then varies with the distance $r$ to the white dwarf, of radius  
R$_{\rm wd}$,
as $n \sim n_0\,(r/R_{\rm wd})^{-2.5}$ (Langer et al. 1982). 
We have set n$_0$, the density at the basis of the column, equal to 
 $1.2\times 10^{15}$ cm$^{-3}$,
corresponding to an accretion rate of $10^{16}$ g\,s$^{-1}$, 
in agreement with the total X-ray luminosity of 6.6\,$10^{32}$ erg\,s$^{-1}$
for a distance of 250 pc (Ramsay et al. 1994)  and assuming 
a typical value  of 10$^{16}$\,cm$^2$ for the polar 
cap surface, and a 0.8 \ms  white dwarf.
The accretion column was approximated by a succession of different slabs 
of constant densities with a maximum extension chosen so that
the density drops by a factor 4 from one slab to the next, up to a maximum 
extension of 50 times the white dwarf radius, i.e. about 1/4 of the 
Roche lobe radius (Mouchet et al. 1997). The lateral extension of each region
was assumed to follow the dipole geometry with a size varying as 
  (r/R$_{\rm wd}$)$^{3/2}$.
The illumination was approximated to be sideways which, taking into account of
the bending of the column, is only true for the furthest parts of the column. 
We have verified however that, in all cases, the lowest slabs of higher 
densities do not contribute significantly to the total flux of the lines 
considered (Mouchet et al. 2002).

The luminosities of the three resonance lines CIV($\lambda$1549), 
NV($\lambda$1240) and OVI($\lambda$1035) were computed with this model, 
first assuming solar abundances.
The results are shown in Fig. 4, where the N/C and O/C ratios are reported. 
Compared to the observed values, these ratios are significantly underestimated 
in this case. Changing the backbody temperature from
50\,eV to 10\,eV reduces significantly the N and O luminosities and does not 
improve the situation (see Fig.\,4).
However, a  hidden very soft X-ray component, contributing significantly
to the ionising flux, could still be present.
For a neutral H column density of $2\times 10^{20}$ cm$^{-2}$, it is found that
a 10\,eV blackbody component  with a luminosity similar to that of the 
bremsstrahlung component, 
could remain undetected and still be compatible with the flux upper 
limit at 100\,\AA{ } of $6\times 10^{-15}$ erg\,cm$^{-2}$\,s$^{-1}$\AA$^{-1}$ 
derived from the EUVE observations. 
Though increasing significantly the N and O luminosities, it still falls 
short of the observed N/C and O/C ratios by factors 10 and 3 respectively. 
We therefore compute next the exact effect of varying the element abundances. 
An N abundance 10 times greater than the solar value was first 
considered. Since parts of the column are optically thick, the N line 
luminosity is not increased proportionally but only by a factor $\sim$4, 
with a corresponding small decrease in the C and O lines but not 
sufficient to reproduce the observed values (see Fig.\,4). 
Assuming our adopted model to be valid, the observed line ratios require 
highly non-solar abundances. 
As the BY Cam line luminosities are somewhat variable, we did not attempt 
here to obtain an exact fit in the abundances but we show in Fig. 4 
a combination where C, N and O are respectively lower by a factor 25,  
higher by 8 and lower by 2, compared to solar values. With these non-solar abundances,
the observed dereddened ratios are well reproduced, within a 
factor 1.01 and 1.22, respectively for NV/CIV and OVI/CIV.
Other combinations may of course be valid and a more detailed modelling 
will be presented elsewhere (Bonnet-Bidaud et al. 2003, in preparation).
These first precise computations of the abundance effects confirm our 
previous suggestions (BM87, Mouchet et al. 1990a) that the peculiar line 
intensities in BY Cam may be due to a conjugated overabundance of N with a 
underabundance of C. 
The strong line of NIII (990\AA){} observed here (as NIV (1718\AA)), 
together with the weaknesses of the CIII 
lines at 977\AA{} and at 1175\AA{}, confirm this. 
The oxygen abundance deviates only slightly  from the solar value.

\begin{figure}
\resizebox{\hsize}{!}
	{\includegraphics[angle=-0,bb=30 170 570 530]{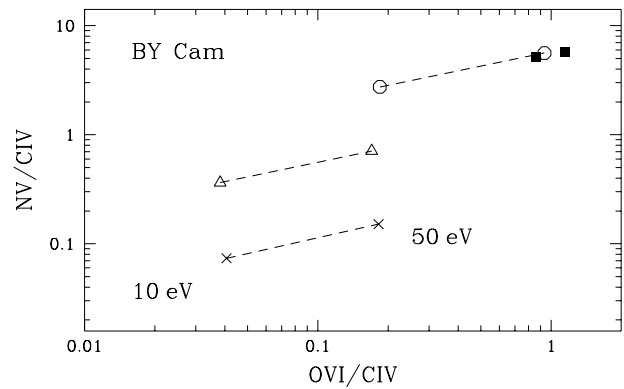}}
\caption{
Predicted broad line ratios computed from CLOUDY plasma code with a simplified 
model of the accretion column (see text). Values for
different ionising spectra with a 20\,keV bremsstrahlung and a blackbody 
temperature ranging from 10\,eV to 50\,eV are shown 
(dashed line) compared to 
the BY Cam observed value (filled squares; left: with no reddening correction, 
right: after reddening correction). Ratios are plotted for solar abundances
(crosses). 
Effects of varying abundances are also shown with a 10 times solar
value for N (open triangles) and a combination where C,N and O are 
respectively 
lower by a factor 25, higher by 8 and lower by 2 compared to solar values 
(open rounds) which fits best the observed value.}
\label{fig4}
\end{figure}

\subsection{Origin of the peculiar CNO composition}
The explanation of such highly non-solar values for some elements 
is still pending.
The nova hypothesis is tempting as the by-product of the H-burning CNO 
cycle is a $^{14}$N production at the expense of $^{12}$C, leading to a CN 
redistribution. But the recent nova V1500 Cyg does not show abnormal
 UV line intensities (Schmidt et al. 1995),
and the ejecta observed in a large sample of novae show enrichment
in all three C, N, O  elements (Livio \& Truran 1994), indicating dredge-up
of matter from the underlying CO core (Starrfield et al. 2000). 
However multicycle models lead to a rearrangement of CNO abundances 
by reprocessing a large quantity of carbon and some fraction of oxygen into 
nitrogen (Kovetz \& Prialnik 1997). 
Enhancement of N and 
depletion of C can indeed be obtained in the context of evolutionary models of 
semi-detached systems in which the secondary surface is polluted by 
re-accretion of material ejected during nova outburts (Marks \& Sarna 1998). 
But the enrichment by a nova-type explosion  requires efficient 
re-accretion of the burnt material as pointed out by Stehle \& Ritter (1999).

As about 1/3 of cataclysmic variables might descend from supersoft X-ray 
binaries (Schenker et al. 2002), the required CNO composition in BY Cam 
can result from steady nuclear burning phases which occur during this 
evolutionary stage.
Interestingly, one of these sources, XMMU J004319.4+411758,  might be 
a progenitor of a magnetic cataclysmic variable (King et al. 2002). 
We note that weak thermonuclear runaways can also have occurred in 
dwarf novae such as U Gem for which the white dwarf 
spectrum reveals a depletion of carbon (Sion et al. 1998) or such as BZ UMa
and EY Cyg
whose UV spectra exhibit a strong NV line and a weak CIV  line 
(Sion 2002, Szkody 2002).

Another possibility for the origin of the peculiar composition 
requires an evolved secondary whose external layers have been stripped off
during episodes of mass loss. Computed evolutionary paths of binary 
systems can produce partially nuclear evolved secondaries prior to contact
(Howell 2001), if the donor star  is sufficiently massive (Haswell et al. 
2002). 
This might explain the pathological CNO  UV line ratios observed in the 
long period system AE\,Aqr (Mouchet et al. 2002), as it has been demonstrated 
by Schenker et al. (2002),  but evolutionary models
adapted for the orbital parameters of BY Cam have still to be computed.

Stronger constraints on the abundance values should be obtained from  
the H-like and He-like lines of C,N, and O which are accessible  with 
XMM-Newton and Chandra. Indeed the first XMM-Newton RGS spectra of 
BY Cam show emission
lines tentatively identified with NVI and NVII lines (Ramsay \& Cropper
2002). Other hydrogen-like and 
helium-like lines of C and O are expected in the RGS range. 
Longer exposed spectra 
are needed to achieve a detailed model of the X-ray emitted plasma 
that would provide a good evaluation of the CNO abundance.
 
\medskip
\it Acknowledgments. \rm We are very grateful to Alain Lecavelier des Etangs 
for his helpful support in FUSE observing preparation and data reduction.

\end{document}